\documentclass[12pt]{iopart}
\usepackage{iopams}

\usepackage[dvipdfmx]{graphicx}
\usepackage{cite}

\begin{document}

\title[AC losses in HTS insert for high field magnet]{Numerical evaluation of AC losses in an HTS insert coil for high field magnet during its energization}

\author{K Kajikawa$^1$, S Awaji$^2$ and K Watanabe$^2$}

\address{$^1$ Research Institute of Superconductor Science and Systems, Kyushu University, Fukuoka 819-0395, Japan}
\address{$^2$ High Field Laboratory for Superconducting Materials, Institute for Material Research, Tohoku University, Sendai 980-8577, Japan}
\ead{kajikawa@sc.kyushu-u.ac.jp}
\begin{abstract}
AC losses in a high temperature superconducting (HTS) insert coil for a 25-T cryogen-free superconducting magnet are numerically calculated during its energization, assuming slab approximation.
The HTS insert coil consists of 68 single pancakes wound with coated conductors and generating a central magnetic field of 11.5~T, in addition to a contribution of 14.0~T from a set of low temperature superconducting (LTS) outsert coils.
Both the HTS and LTS coils are cooled using cryocoolers, and energized simultaneously up to 25.5~T with a constant ramp rate for 60~min.
The influences of the magnitudes and orientations of the locally applied magnetic fields, magnetic interactions between turns, and the transport currents flowing in the windings are taken into account in the AC loss calculations.
The locally applied fields are separated into axial and radial components, and the individual contributions of these field components to the AC losses are summed simply to obtain the total losses.
The contribution of the axial field component to the total AC loss is large at the early stages of the energization, whereas the total losses monotonically increase with time after the contribution of the radial field component becomes sufficiently large.
\end{abstract}

\maketitle

\section{Introduction}

High temperature superconducting (HTS) wires with long lengths composed of Y-based or rare-earth-based superconductors have been developed and have recently become commercially available.
These HTS wires, called coated conductors, are in the form of a tape with a width of several millimetres and include a superconducting (SC) layer with a thickness of a few micrometers.
Since the aspect ratio of the cross section of the SC layer, which is defined as the ratio of the width to the thickness, is more than 1000, and also, these superconductors themselves have a layered crystal structure, the electromagnetic properties such as critical current density and AC loss have an anisotropy and depend on the direction of an externally applied magnetic field, as well as its magnitude.
HTS magnets wound with coated conductors are also expected to be operated, not only by immersion cooling with liquid helium or nitrogen, but also by conduction cooling with a cryocooler.
Therefore, evaluation of the AC losses during the charging up or down of the HTS magnets cooled using cryocoolers has become very important in order to prevent thermal runaway of the magnets.
This is because the thermal runaway is caused by excess heating power that is beyond the cooling power of the cryocoolers.
It has also become well known that the transport current flowing in the coated conductor itself affects the AC loss property in an externally applied cyclic magnetic field~\cite{Amemiya,Mawatari07,Kajikawa_PC10}.
Currently, there is no useful expression to evaluate the magnitude of AC loss in coated conductors that are exposed to both external magnetic fields and transport currents varying at constant sweep rates.

A set of recent reports have estimated AC losses in HTS insert coils wound with coated conductors for a 32-T high field magnet~\cite{Gavrilin,Lu}.
Since these HTS inserts are immersed in liquid helium, the total amount of AC losses during the charging or discharging process is focused there.
In the case of an HTS insert coil cooled using a cryocooler, however, it may be necessary to evaluate the magnitude of the instantaneous AC loss.
The proposed expression to estimate the AC losses in the HTS inserts for the 32-T magnet is also applicable to the case where the radial component of a local magnetic field, applied to every part of the windings in solenoid coils, is larger than the full penetration field.
However, such a situation cannot be realized for an HTS insert in a high field magnet with a relatively large height.
It is well known that, if the magnitude of the local magnetic field does not exceed the full penetration field, the magnetic interactions between the SC wires strongly affect the AC loss properties, for example, for coils wound with NbTi wires~\cite{Zenkevitch,Sumiyoshi,Kajikawa_IPCS167}, stacks of BSCCO tapes~\cite{Suenaga,Kajikawa_IEEJ01}, and stacks of coated conductors~\cite{Iwakuma,Grilli06}.
Furthermore, although numerical analyses of AC loss in bundles of coated conductors have been carried out for their stacks~\cite{Grilli06,Yuan,Kajikawa_PC09,Prigozhin}, single pancake coils~\cite{Grilli07,Pardo}, racetrack coils~\cite{Zermeno}, and power transmission cables~\cite{Sato,Kajikawa_PC06}, the numbers of coated conductors used for numerical modelling are limited to less than 200.

In this paper, AC losses in an HTS insert coil designed for a high field magnet~\cite{Awaji_IEEE-TAS} are numerically evaluated.
It is assumed that the HTS insert is composed of a stack of single pancake coils wound using coated conductors and energized up to a rated current at a constant ramping rate under conduction cooling using cryocoolers.
The influences of the magnitudes and orientations of locally applied magnetic fields, magnetic interactions between turns, and transport currents flowing in the windings are taken into account in the calculations of the AC losses.

\section{Specifications of the HTS insert for the high field magnet}

\begin{figure}[tbp]
\centering
\begin{tabular}{cc}
\includegraphics*[scale=1.0]{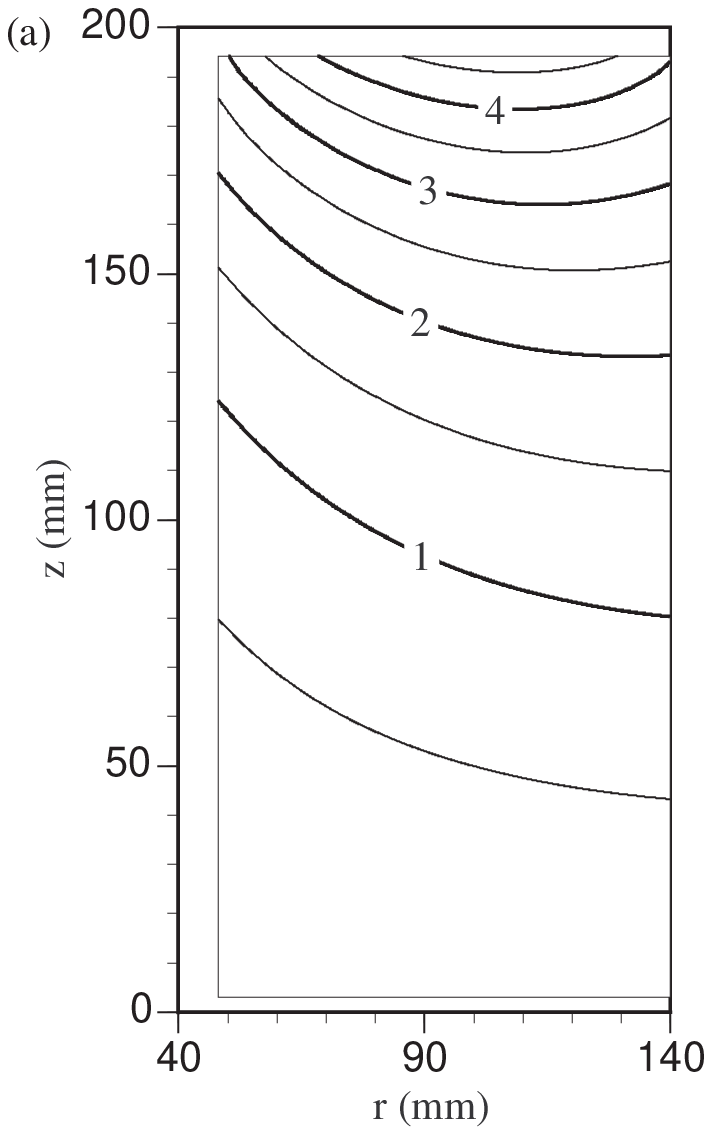}&
\includegraphics*[scale=1.0]{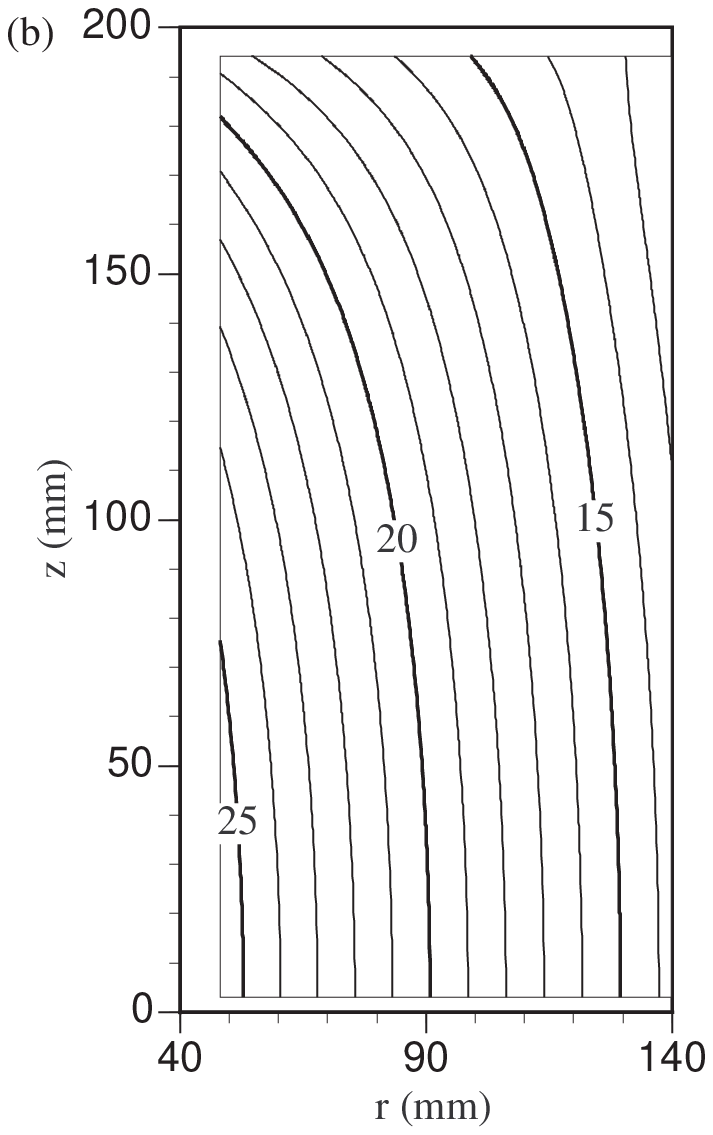}\\
\includegraphics*[scale=1.0]{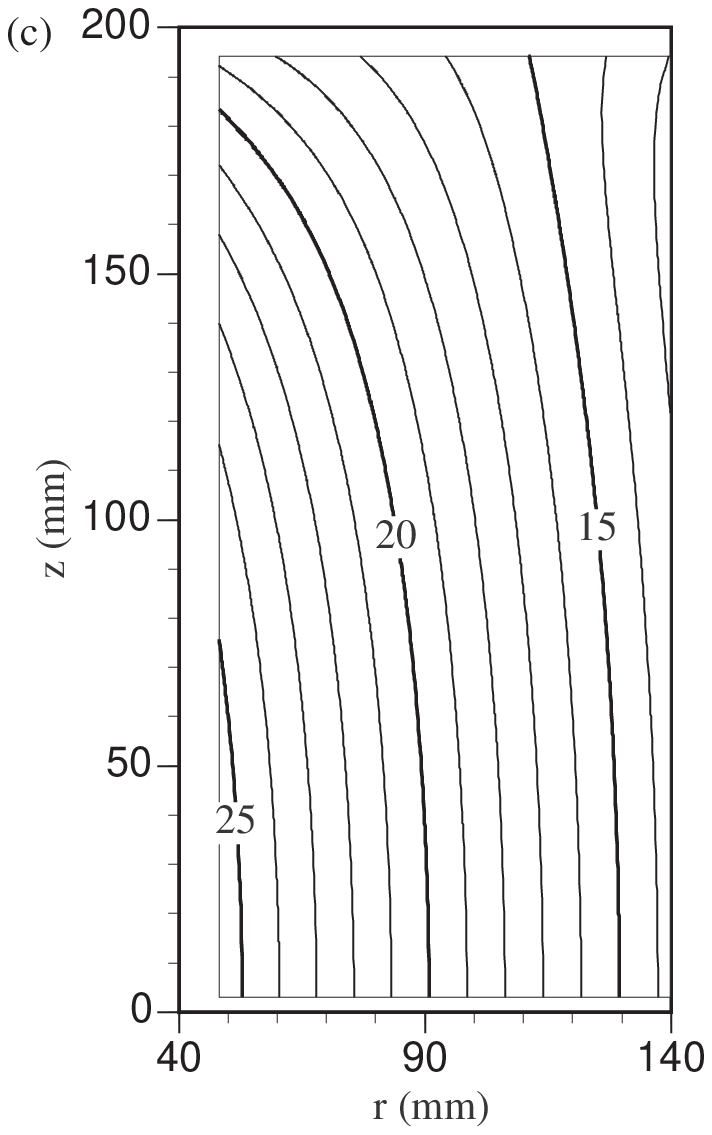}&
\includegraphics*[scale=1.0]{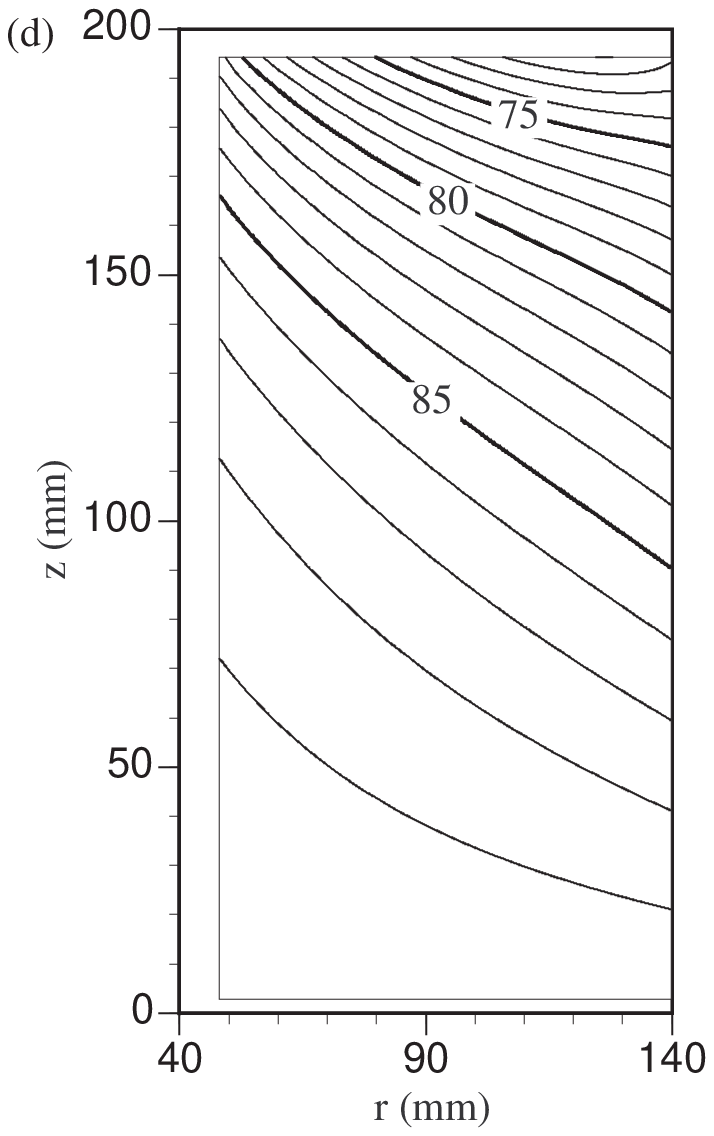}\\
\end{tabular}
\caption{Contour maps of (a) radial field $B_r\!$~(T), (b) axial field $B_z\!$~(T), (c) magnitude $\left|B\right|\!$~(T), and (d) angle $\theta\!$~(deg.) inside upper half of HTS insert coil with LTS outsert coils.
All the coils generate a total central field of 25.5~T.}
\label{fig1}
\end{figure}

A 25-T-class cryogen-free SC magnet has recently been designed~\cite{Awaji_IEEE-TAS}.
An HTS insert coil is located coaxially inside six low temperature superconducting (LTS) outsert coils.
The combination of LTS coils wound using NbTi or Nb$_3$Sn wire generates a central magnetic field of 14.0~T.
The specifications of the HTS insert are listed in \tref{tbl1}.
Gd-based coated conductors with a width of 5.00~mm and a thickness of 0.13~mm are wound into 68 single pancakes with 438 turns.
The inner radius, outer radius, and height of the HTS insert are 48.0~mm, 140~mm, and 394.4~mm, respectively.
This HTS insert is operated at 135~A and generates a central magnetic field of 11.5~T.
Therefore, the total central field generated by the combination of the HTS insert and the LTS coils becomes 25.5~T.
The HTS insert and the LTS coils are cooled individually using cryocoolers.
Two 2-stage Gifford-McMahon cryocoolers with 1.5~W at 4.2~K are used for the cooling of the HTS insert.
The energizing time of up to 25.5~T is set to less than 60~min.
\Fref{fig1} shows the numerical results of profiles of magnetic fields inside the upper half of the HTS insert coil combined with the LTS outsert coils.
All the coils generate a total central field of 25.5~T.
Figures \ref{fig1}(a), (b), (c), and (d) plot the contour maps of the radial, $B_r\!$~(T), and axial, $B_z\!$~(T), components of the local magnetic field, its magnitude $\left|B\right|\!$~(T), and angle $\theta\!$~(deg.), respectively.
\begin{table}[tbp]
\caption{\label{tbl1}Specifications of HTS insert for high field magnet~\cite{Awaji_IEEE-TAS}.}
\begin{indented}
\item[]\begin{tabular}{@{}lc}
\br
Parameter&Value\\
\mr
Material&GdBa$_2$Cu$_3$O$_y$\\
Wire size&5.00~mm$^w$ $\times$ 0.13~mm$^t$\\
Operating current&135~A\\
Inner radius&48.0~mm\\
Outer radius&140~mm\\
Coil height&394.4~mm\\
Number of single pancakes&68\\
Number of turns per single pancake&438\\
Central field&11.5~T\\
\br
\end{tabular}
\end{indented}
\end{table}

The magnetization losses, $W_{\mathrm{magnetization}}$, per unit volume per AC cycle in stacked BSCCO tapes exposed to the external transverse magnetic fields, $\boldsymbol{B}=\left(B_x,B_y,0\right)$, with arbitrary angles have been evaluated experimentally using the following relationship~\cite{Fukuda}
\begin{eqnarray}
W_{\mathrm{magnetization}}\!\left(\boldsymbol{B}\right)&={}-\!\oint\!\boldsymbol{M}\!\cdot\;\!\!\rmd\boldsymbol{B}\nonumber\\
&={}-\!\oint\!M_y\!\left(\boldsymbol{B}\right)\rmd B_y-\!\oint\!M_x\!\left(\boldsymbol{B}\right)\rmd B_x\nonumber\\
&\simeq{}-\!\oint\!M_y\!\left(B_y\right)\rmd B_y-\!\oint\!M_x\!\left(B_x\right)\rmd B_x\nonumber\\
&=W_{\mathrm{parallel}}\!\left(B_y\right)+W_{\mathrm{perpendicular}}\!\left(B_x\right)\label{eqn:loss1}
\end{eqnarray}
where the $y$- and $x$-directions are set parallel and perpendicular to the wide surfaces of the tapes with relatively long lengths in the $z$-direction, respectively, and $\boldsymbol{M}=\left(M_x,M_y,0\right)$ represents the magnetization of the stacked tapes.
It can be seen in \eref{eqn:loss1} that the inner product of the two vectors $\boldsymbol{M}$ and $\rmd\boldsymbol{B}$, which is used to evaluate the area of the magnetization loop, is deformed into the sum of two components, $M_y\,\rmd B_y$ and $M_x\,\rmd B_x$.
Although the components $M_y$ and $M_x$ of the magnetization are generally given as a function of the local magnetic field, $\boldsymbol{B}$, the first and second terms in \eref{eqn:loss1} could correspond to the contributions of the parallel and perpendicular components, respectively, if it is assumed that $M_y$ and $M_x$ depend only on $B_y$ and $B_x$, respectively.
Therefore, $W_{\mathrm{magnetization}}$ could be given by the simple sum of the individual contributions of the parallel-field loss, $W_{\mathrm{parallel}}$, (due to the parallel field $B_y$) and a perpendicular-field loss, $W_{\mathrm{perpendicular}}$, (due to the perpendicular field $B_x$).

By extending the above-mentioned idea to evaluate AC losses in a solenoid magnet composed of pancake coils wound using coated conductors, the total loss, $Q_{\mathrm{total}}$, in the magnet can be expressed by
\begin{eqnarray}
&Q_{\mathrm{total}}=\sum_{k=1}^{N\!P}Q_k\label{eqn:loss2}\\
&Q_k\!\left(\boldsymbol{B},I\right)=Q_{\mathrm{parallel}}\!\left(B_z,I\right)+Q_{\mathrm{perpendicular}}\!\left(B_r,I\right)\label{eqn:loss3}
\end{eqnarray}
where $P$ is the number of single pancakes and $N$ is the number of turns per single pancake.
\Eref{eqn:loss2} means that the AC losses in the individual turns are estimated and summed.
The local magnetic field, $\boldsymbol{B}$, exposed to one turn under consideration, has axial and radial components $B_z$ and $B_r$, respectively, which are directed parallel and perpendicular to the wide surface of the HTS tape wound flatwise, respectively.
Thus, \eref{eqn:loss3} means that the AC loss of the $k$-th turn can be obtained simply by adding the parallel-field loss, $Q_{\mathrm{parallel}}$, determined by the axial field, $B_z$, and the perpendicular-field loss, $Q_{\mathrm{perpendicular}}$, determined by the radial field, $B_r$.
However, the influence of a transport current, $I$, on the losses has to be taken into account.
In order to estimate $Q_{\mathrm{parallel}}$ in the $k$-th turn, the SC layer of the coated conductor can be considered as an infinite slab with a width equal to the thickness, $d$, of the SC layer because $d$ is much smaller than the tape width, $2a$.
On the other hand, the magnetic interactions between turns~\cite{Kajikawa_IEEJ01,Kajikawa_PC09} have to be taken into account in order to estimate $Q_{\mathrm{perpendicular}}$, as discussed in \sref{sec:magnetic_interaction}.

In order to estimate the AC losses of pancake coils using \eref{eqn:loss2} and \eref{eqn:loss3}, the dependence of the critical current density, $J_c$, on the local magnetic field has to be taken into account.
Critical current densities of a Gd-based coated conductor as a function of the magnitudes, $\left|B\right|$, and the angles, $\theta$, of externally applied magnetic fields have been measured experimentally at a fixed temperature of 4.2~K~\cite{Awaji_CSSJ}.
In this study, these experimental results are approximated by means of the least-square technique in the range of $0^{\circ}\le\theta\le90^{\circ}$ using the equation
\begin{equation}
J_c\!\left(\left|B\right|\!,\theta\right)=\frac{\alpha\!\left|B\right|^{-\Gamma}}{\sqrt{\cos^2\!\left[\displaystyle\frac{\pi}{2}\!\left(\!\frac{\theta}{90^{\circ}}\!\right)^{\!p}\,\right]+\displaystyle\frac{1}{\gamma^2}\sin^2\!\left[\displaystyle\frac{\pi}{2}\!\left(\!\frac{\theta}{90^{\circ}}\!\right)^{\!q}\,\right]}}
\end{equation}
where the angle $\theta=0^{\circ}$ means that the external field is perpendicular to the wide surface of the tape, whereas the angle $\theta=90^{\circ}$ means that the field is parallel to the wide surface.
As a result, a set of fitting parameters $\left(\alpha,\Gamma,\gamma,p,q\right)$ is obtained for $J_c\!$~(A/m$^2$), where $\alpha=2.05\times10^{11}$, $\Gamma=0.682$, $\gamma=7$, $p=1.86$, and $q=12.1$.

\section{Influence of magnetic interaction between tapes on AC losses}
\label{sec:magnetic_interaction}

In order to evaluate the influence of magnetic interaction between SC tapes, AC losses in thin strips are numerically calculated by means of a one-dimensional finite element method formulated using only a current vector potential, $T$~\cite{Sato,Kajikawa_PC06,Hashizume,Ichiki}.
Let us consider SC strips with an infinite length in the $Z$-direction, of which the thickness, $d$, in the $Y$-direction is much smaller than the width, $2a$, in the $X$-direction.
The $N$ strips are also stacked face-to-face at even intervals of $g$ in the $Y$direction, and one of the strips under consideration is exposed to an external magnetic field, $H_{eY}$, in the $Y$-direction.
In this case, a local current density, $\boldsymbol{J}$, in the strip under consideration has only the $Z$-component $J\!\left(X\right)$ as a function of the position $X$, and the current vector potential, $\boldsymbol{T}$, defined by $\boldsymbol{J}=\nabla\times\boldsymbol{T}$, has only the $Y$-component $T\!\left(X\right)$.
Hence, the relationship between $J$ and $T$ is given by $J=\partial T/\partial\!X$.
A governing equation is expressed by Faraday's law formulated with the potential $T$~\cite{Sato,Kajikawa_PC06}
\begin{equation}
\frac{\partial}{\partial\!X}\!\left(\!\rho\frac{\partial T}{\partial\!X}\!\right)=\mu_0d\,\frac{\partial}{\partial t}\sum_{k=1}^{N}\!\int_{C_k^{\prime}}\!\!\!F\!\left(X,X^{\prime}\right)\!\frac{\partial T}{\partial\!X^{\prime}}\,\rmd X^{\prime}+\mu_0\frac{\partial\;\!\!H_{eY}}{\partial t}\label{eq:governing_equation}
\end{equation}
where $\rho$ is the electrical resistivity of strip, $F$ is the geometric factor determined by the strip arrangement, and $C_k^{\prime}$ is the line path along the $k$-th strip.
The second term on the right-hand side of \eref{eq:governing_equation} represents the time derivative of the external field.
On the other hand, the first term is the time derivative of the total magnetic fields generated by the currents induced in all the strips.
\Eref{eq:governing_equation} is discretized by means of the Galerkin method for space and the backward difference method for time.
The boundary conditions are given by
\begin{equation}
\left\{T\!\left(a\right)-T\!\left(-a\right)\right\}\;\!\!d=I_k
\end{equation}
where $I_k$ is the transport current in the $k$-th strip and $X=\pm a$ are both edges of the strips.

Bean's critical state model including the flux-flow state is used here for the relationship between an electric field, $E$, and the current density, $J$, as follows~\cite{Bean,Kajikawa_IEEE-TAS03}
\begin{equation}
E\!\left(J\right)=\cases{0&for $\left|J\right|\le J_c$\\
\rho_f\!\left(\left|J\right|-J_c\right)\!\frac{J}{\left|J\right|}&for $\left|J\right|>J_c$\\}
\end{equation}
where $\rho_f$ is the flux-flow resistivity.
Therefore, $\rho$ is given by
\begin{equation}
\rho\!\left(J\right)=\frac{E\!\left(J\right)}{J}=\cases{0&for $\left|J\right|\le J_c$\\
\rho_f\!\left(1-\frac{J_c}{\left|J\right|}\right)&for $\left|J\right|>J_c$\\}
\end{equation}
Since $\rho$ depends on $T$ through $J$ and is nonlinear, the Newton-Raphson method is used for iterative calculations.
The AC loss per unit volume per AC cycle, $W$, can be obtained by
\begin{equation}
W=\frac{1}{2a}\!\oint\!\rmd t\!\int_{-a}^{a}\!\rho J^2\,\rmd X=\frac{1}{2a}\!\oint\!\rmd t\!\int_{-a}^{a}\!\rho\!\left(\!\frac{\partial T}{\partial\!X}\!\right)^{\!\!2}\!\rmd X
\end{equation}

The parameters for numerical calculations using the finite element method are listed in \tref{tbl2}.
The width, $2a$, and thickness, $d$, of the SC strips are 5~mm and 2~$\mu$m, respectively.
The number, $N$, of strips stacked at even intervals, $g=0.21$~mm, are varied up to a maximum of $N=438$.
The tape width, $2a$, of each strip is discretized into 100 line elements with identical lengths.
$J_c$ is fixed at 2~MA/cm$^2$, and $\rho_f$ is assumed to be 10~$\mu\Omega$cm~\cite{Kiss}.
The cyclic AC magnetic fields with amplitudes $H_m$ are applied perpendicularly to the stacked strips.
The frequency is fixed at 1~Hz.
\begin{table}[tbp]
\caption{\label{tbl2}Parameters for numerical calculations using finite element method.}
\begin{indented}
\item[]\begin{tabular}{@{}lc}
\br
Parameter&Value\\
\mr
Tape width, $2a$&5~mm\\
Thickness of SC strip, $d$&2~$\mu$m\\
Spatial interval between strips, $g$&0.21~mm\\
Number of stacked strips, $N$&1 to 438\\
Number of elements with identical length for each strip&100\\
Critical current density, $J_c$&2~MA/cm$^2$\\
Flux-flow resistivity, $\rho_f$&10~$\mu\Omega$cm\\
AC frequency&1~Hz\\
\br
\end{tabular}
\end{indented}
\end{table}
\begin{figure}[tbp]
\centering
\includegraphics*[scale=1.0]{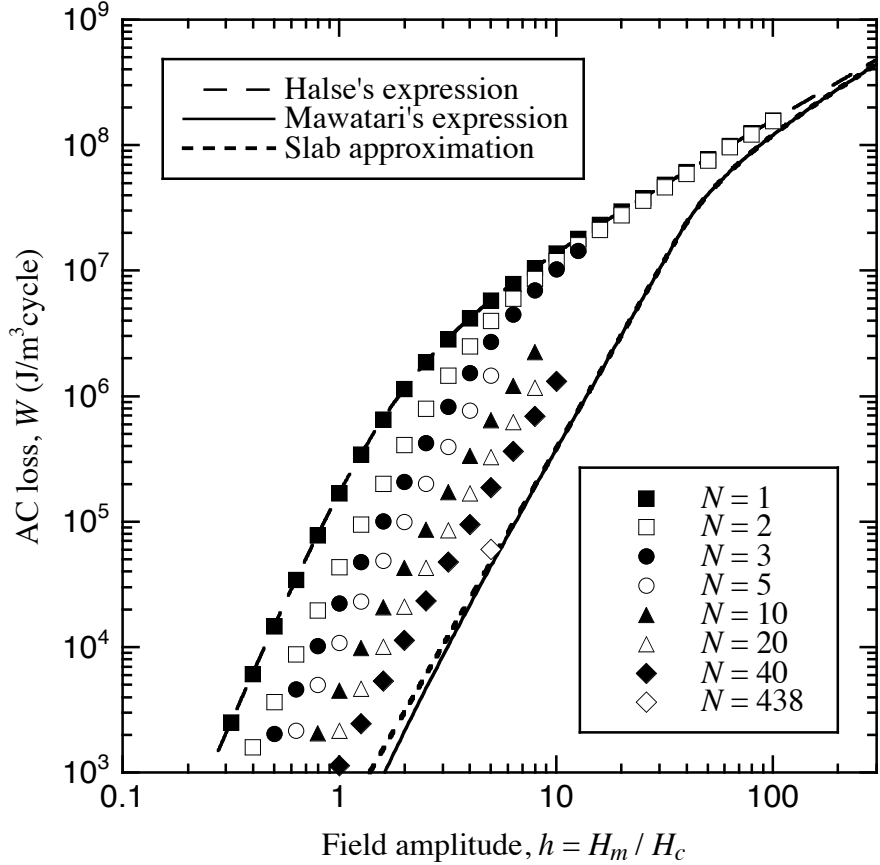}
\caption{Comparison between numerical and theoretical results for AC losses of stacked strips.}
\label{fig2}
\end{figure}

\Fref{fig2} shows a comparison between the numerical and theoretical results for the AC losses of the stacked strips.
The broken line in \fref{fig2} represents a theoretical curve for a single strip obtained originally by Halse~\cite{Halse}, which can be expressed as~\cite{Brandt,Zeldov}
\begin{equation}
W=W_0\!\left(2\ln\cosh h-h\tanh h\right)
\end{equation}
where $W_0=2\mu_0J_caH_c$ and $h=H_m/H_c$, with $H_c=J_cd/\pi$.
The solid line in \fref{fig2} is drawn using a theoretical expression for infinitely stacked strips obtained by Mawatari~\cite{Mawatari96}, which can be expressed as
\begin{equation}
W=\frac{W_0}{c^2}\!\!\int_{0}^{h}\!\!\left(h-2\xi\right)\ln\!\left(1+\frac{\sinh^2\!c}{\cosh^2\!\xi}\right)\!\rmd\xi
\end{equation}
where $c=\pi a/g$.
The dashed line in \fref{fig2} is for an infinite slab with a width of $2a$ given by~\cite{Kajikawa_PC09,London,Clem}
\begin{equation}
W=\frac{1}{\lambda}\times\cases{\frac{2\mu_0}{3}\frac{H_m^3}{H_p}&for $H_m\le H_p$\\
2\mu_0H_pH_m\!\left(1-\frac{2}{3}\frac{H_p}{H_m}\right)&for $H_m>H_p$\\}
\end{equation}
where $\lambda=d/g$ and $H_p=\lambda J_ca$.
Let us first compare these three theoretical curves with one another.
The AC losses in the infinitely stacked strips can be well-explained by the slab approximation, apart from a range of very small amplitudes.
This is because the spatial interval, $g$, between the strips is more than 20 times shorter than the tape width, $2a$.
Also, there is no discrepancy between the theoretical results in a range of very large amplitudes.
On the other hand, if the amplitude becomes small, the AC losses in the infinitely stacked strips are much smaller than those for the single strip, because of the magnetic interaction between the strips.

\begin{figure}[tbp]
\centering
\includegraphics*[scale=1.0]{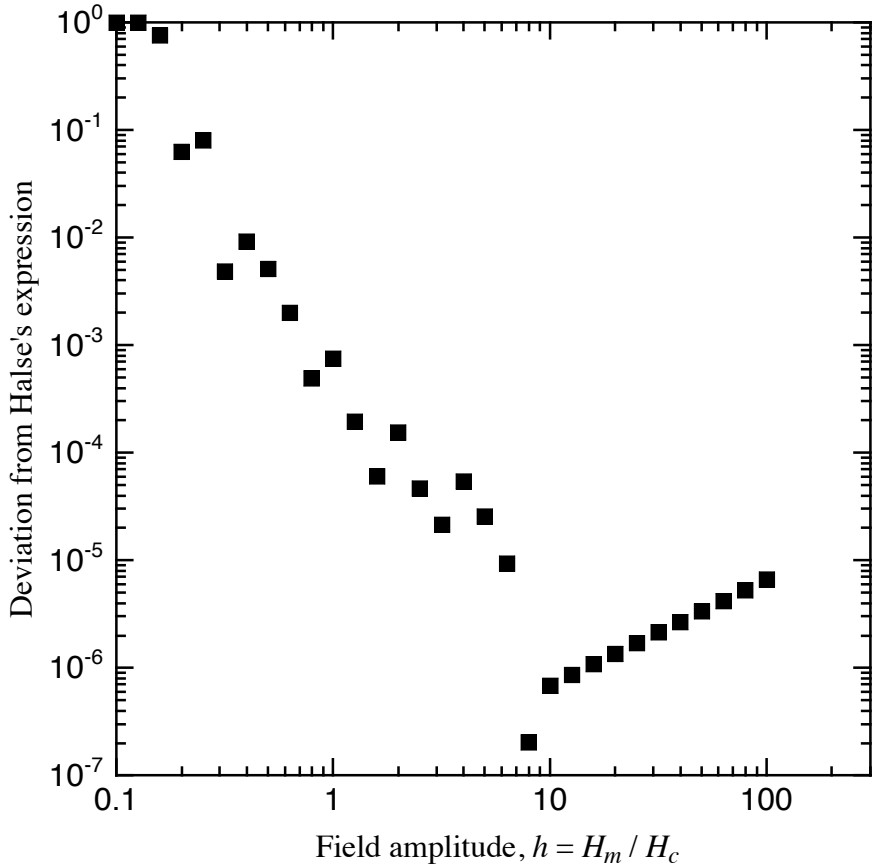}
\caption{Deviation of numerical results for single strip from Halse's expression~\cite{Halse,Brandt,Zeldov}.}
\label{fig3}
\end{figure}

The symbols in \fref{fig2} represent the numerical results for the AC losses calculated by means of the finite element method.
It is found that the numerical results for the single strip have good agreement with Halse's expression.
\Fref{fig3} plots the deviation of the numerical results for a single strip from Halse's expression.
It can be seen that the numerical errors are less than 1\% if the normalized field amplitudes, $h$, are larger than 0.3.
This means that the theoretical values can be reproduced accurately by numerical calculations, except for cases of small amplitudes.
If the number of elements is increased to more than 100, the numerical errors could be reduced in a wider range of field amplitudes.
$N$ is increased up to a maximum of 438 in \fref{fig2}.
It is found that the AC losses decrease with increasing strip numbers due to the magnetic interaction between the strips, and the AC loss for 438 strips asymptotically approaches the theoretical values for both the infinite stack and slab.

\begin{figure}[tbp]
\centering
\includegraphics*[scale=1.0]{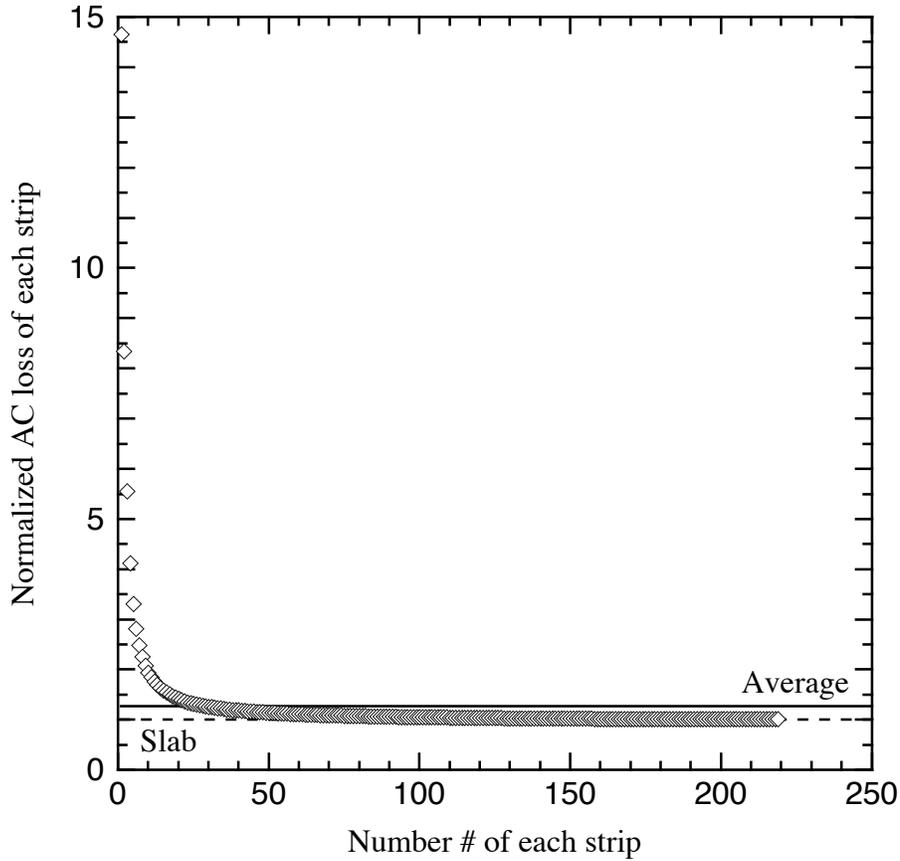}
\caption{AC loss for each strip within the stack of 438 strips.
In the interests of symmetry, only the AC losses for half of the stack are calculated and plotted.}
\label{fig4}
\end{figure}

\Fref{fig4} shows the AC loss for each strip within the stack of 438 strips.
$h$ is fixed at 5 in a range of small amplitudes.
In the interests of symmetry, only the AC losses for half of the stack are calculated and plotted here.
The strip labelled \#1 is located at the edge of the stack, and strip \#219 is at the centre of the stack.
The AC losses are also normalized by those for the infinite slab, which corresponds to the dashed line in \fref{fig4}.
A solid line in \fref{fig4} represents an averaged value of AC losses for the 438 strips.
It can be seen that the AC losses in only several strips close to the edge are larger than those of the others.
On the other hand, the AC losses for almost all the strips in the central part are also very close to that of the infinite slab.
Therefore, the averaged value of AC losses for all the strips shows reasonably good agreement with that of the infinite slab.
This means that the AC losses in the stacked strips can be estimated with relatively small errors on the basis of the slab approximation, if the number of strips is large enough and the spatial intervals between the strips are short.

\section{Theoretical expressions of AC losses}

\begin{figure}[tbp]
\centering
\includegraphics*[scale=1.0]{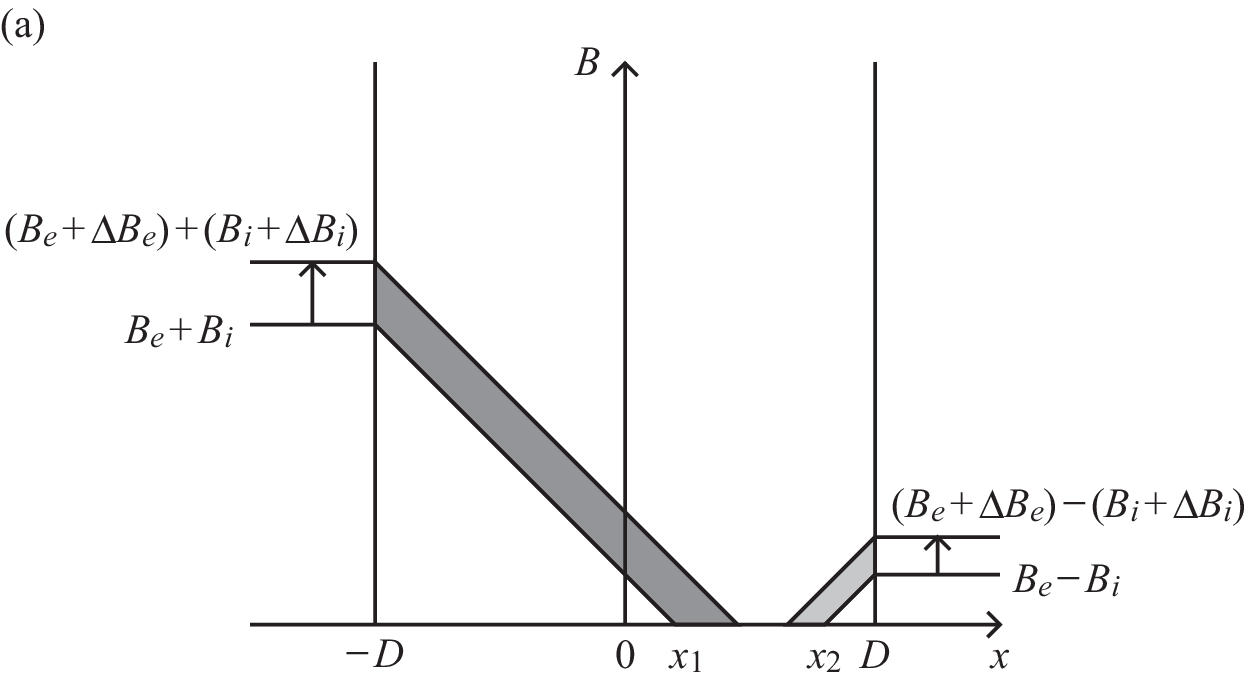}\\
\includegraphics*[scale=1.0]{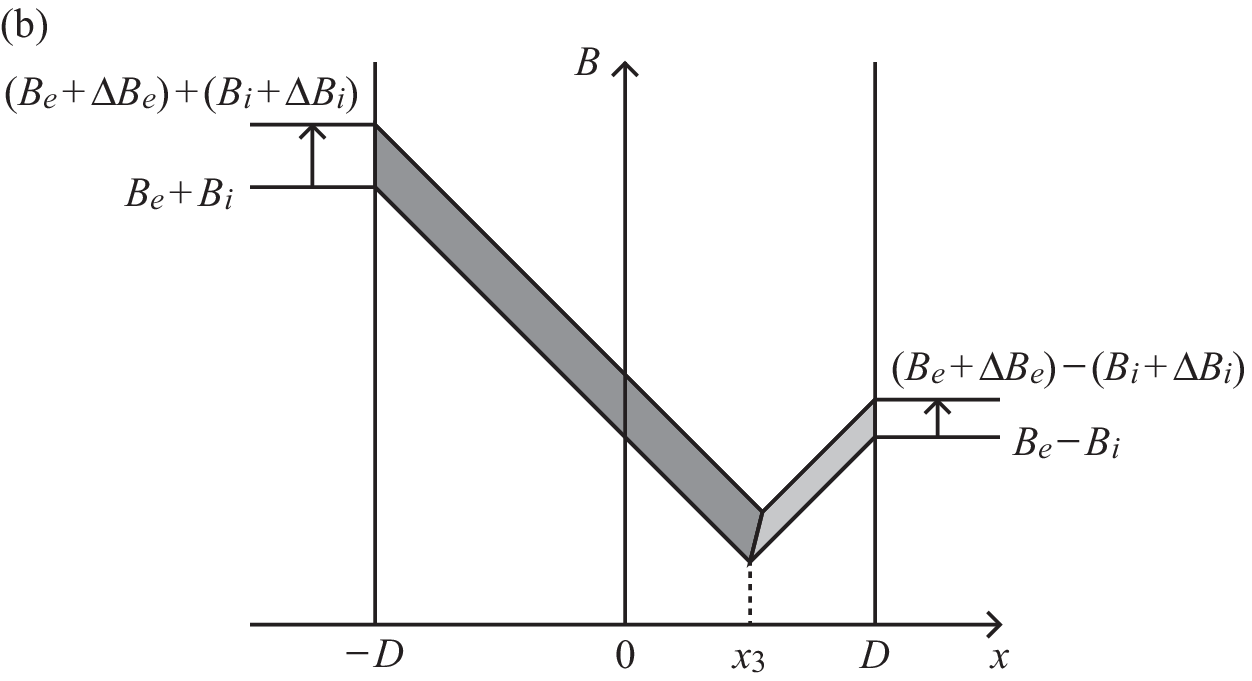}
\caption{Profiles of magnetic fields inside the infinite slab with transport current in external magnetic field for (a) $B_e<B_p$ and (b) $B_e\ge B_p$.}
\label{fig5}
\end{figure}

Let us derive theoretical expressions for the AC losses in an infinite slab with a width of $2D$, as shown in \fref{fig5}.
The Bean model~\cite{Bean}, in which the critical current density, $J_c$, is independent of the magnitude of the local magnetic field, is assumed here.
An external magnetic field, $B_e$, is applied to the infinite slab with a transport current $I$, which generates a self-field, $B_i$, given by $B_i=\left(I/I_c\right)\!B_p$, with critical current $I_c$.
The field profile for the case where $B_e$ is less than the full penetration field, $B_p\left(=\mu_0J_cD\right)$, is shown in \fref{fig5}(a), whereas \fref{fig5}(b) is the case where $B_e\ge B_p$.
The flux front positions for the field profiles, $x_1$, $x_2$, and $x_3$, in \fref{fig5} are given by
\numparts
\begin{eqnarray}
x_1&=\frac{B_e+B_i-B_p}{B_p}D\\
x_2&=\frac{B_p-\left|B_e-B_i\right|}{B_p}D\\
x_3&=\frac{B_i}{B_p}D
\end{eqnarray}
\endnumparts

If $B_e$ and $I$ are incremented by $\Delta B_e$ and $\Delta I$, respectively, during a period $\Delta t$, and the corresponding increment of the self-field is $\Delta B_i=\left(\Delta I/I_c\right)\!B_p$, the distribution of the induced local electric field, $E$, can be expressed as
\begin{equation}
E\!\left(x\right)=\cases{\frac{\left(\Delta B_e+\Delta B_i\right)\!\left(x_1-x\right)}{\Delta t}&for $B_e<B_p$, $-D\le x\le x_1$\\
0&for $B_e<B_p$, $x_1<x\le x_2$\\
-\frac{\left(\Delta B_e-\Delta B_i\right)\!\left(x-x_2\right)}{\Delta t}&for $B_e<B_p$, $x_2<x\le D$\\
\frac{\left(\Delta B_e+\Delta B_i\right)\!\left(x_3-x\right)}{\Delta t}&for $B_e\ge B_p$, $-D\le x\le x_3$\\
-\frac{\left(\Delta B_e-\Delta B_i\right)\!\left(x-x_3\right)}{\Delta t}&for $B_e\ge B_p$, $x_3<x\le D$\\}
\end{equation}
On the other hand, the distribution of the local current density, $J$, is given by
\begin{equation}
J\!\left(x\right)=\cases{J_c&for $B_e<B_p$, $-D\le x\le x_1$\\
0&for $B_e<B_p$, $x_1<x\le x_2$\\
-\frac{B_e-B_i}{\left|B_e-B_i\right|}J_c&for $B_e<B_p$, $x_2<x\le D$\\
J_c&for $B_e\ge B_p$, $-D\le x\le x_3$\\
-\frac{B_e-B_i}{\left|B_e-B_i\right|}J_c&for $B_e\ge B_p$, $x_3<x\le D$\\}
\end{equation}
Therefore, the AC loss power $\dot{Q}$ per unit volume during the energization, $\dot{B_e}\ge 0$ and $\dot{I}\ge 0$, can be obtained as
\begin{eqnarray}
&\dot{Q}=\frac{1}{2D}\lim_{\Delta t\to 0}\int_{-D}^{D}\!E\!\left(x\right)\!J\!\left(x\right)\rmd x\nonumber\\
&=\frac{B_p^2}{2\mu_0}\times\cases{\frac{B_e^2}{B_p^2}\frac{\dot{B}_e}{B_p}+\!\left[\!\left(2\,\,\!\frac{B_e}{B_p}\frac{I}{I_c}-\frac{B_e^2}{B_p^2}\right)\!\frac{\dot{B}_e}{B_p}\right.&\\
\left.\qquad\qquad\quad\!{}+\!\left(\frac{B_e^2}{B_p^2}+\frac{I^2}{I_c^2}\right)\!\frac{\dot{I}}{I_c}\right]&for $0\le B_e<\displaystyle\frac{I}{I_c}B_p$\\
\frac{B_e^2}{B_p^2}\frac{\dot{B}_e}{B_p}+\!\left(\frac{I^2}{I_c^2}\frac{\dot{B}_e}{B_p}+2\,\,\!\frac{B_e}{B_p}\frac{I}{I_c}\frac{\dot{I}}{I_c}\right)&for $\displaystyle\frac{I}{I_c}B_p\le B_e<B_p$\\
\frac{\dot{B}_e}{B_p}+\!\left(\frac{I^2}{I_c^2}\frac{\dot{B}_e}{B_p}+2\,\,\!\frac{I}{I_c}\frac{\dot{I}}{I_c}\right)&for $B_e\ge B_p$\\}\label{eqn:loss4}
\end{eqnarray}
where $\dot{B_e}$ and $\dot{I}$ represent the time derivatives of $B_e$ and $I$, respectively.
The first terms on the right-hand side of \eref{eqn:loss4} represent the contribution from the external magnetic field in the case without a transport current, whereas the second terms arise due to the existence of the transport current.
The theoretical expressions \eref{eqn:loss4} are used to estimate both the parallel- and perpendicular-field losses for the HTS insert in the next section.

\section{AC loss calculations of HTS insert}

The numerical parameters for AC loss calculations of the HTS insert are summarized in \tref{tbl3}.
Almost all the parameters are based on the designed HTS insert~\cite{Awaji_IEEE-TAS} listed in \tref{tbl1}.
The LTS coils and HTS insert are simultaneously energized up to 25.5~T in 60~min.
\begin{table}[tbp]
\caption{\label{tbl3}Numerical parameters for AC loss calculations of HTS insert.}
\begin{indented}
\item[]\begin{tabular}{@{}lc}
\br
Parameter&Value\\
\mr
Tape width, $2a$&5~mm\\
Thickness of SC layer, $d$&2~$\mu$m\\
Winding pitch of HTS tape, $g$&0.21~mm\\
Gap length between pancakes&0.8~mm\\
Average radius of innermost turn&48.13~mm\\
Average radius of outermost turn&139.90~mm\\
Distance between centres of top and bottom pancakes&388.6~mm\\
Number of single pancakes, $P$&68\\
Number of turns per single pancake, $N$&438\\
Operating current&135~A\\
Energizing time to 25.5~T&60~min\\
Energizations of LTS coils \& HTS insert&Simultaneous\\
\br
\end{tabular}
\end{indented}
\end{table}

\begin{figure}[tbp]
\centering
\includegraphics*[scale=1.0]{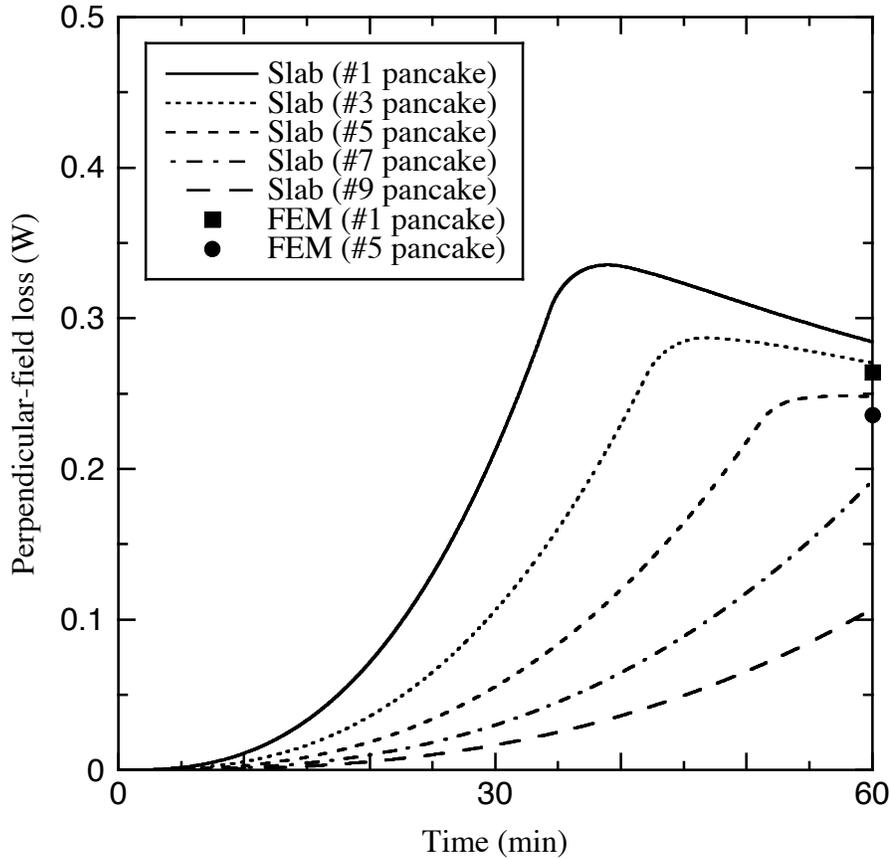}
\caption{Time evolution of perpendicular-field losses in selected pancake coils of HTS insert.
The curves are obtained using the slab approximation, whereas the symbols show the numerical results determined using the finite element method.}
\label{fig6}
\end{figure}

\Fref{fig6} shows the time evolution of AC losses during energization for the radial components, $B_r$, of the magnetic fields perpendicular to the HTS tapes in some selected pancake coils of the HTS insert.
The numbering for the pancakes is performed in series from the top to the bottom of the HTS insert.
The curves in \fref{fig6} are obtained using the theoretical expressions \eref{eqn:loss4} for an infinite slab with a width equal to the tape width $2a$, whereas the symbols show the numerical results obtained by means of the finite element method (FEM), formulated with the potential $T$.
In the FEM calculations, the number, $N$, of stacked strips, each of which is modelled as 50 line elements of identical lengths, is fixed at 438, and no transport current is applied.
However, different values of $J_c$ and $B_r$ are used for each strip within the stack.
It can be seen that the perpendicular-field losses in pancake \#1 are the largest.
It is also found that both the results from the slab approximation and the FEM show reasonably good agreement.

\begin{figure}[tbp]
\centering
\includegraphics*[scale=1.0]{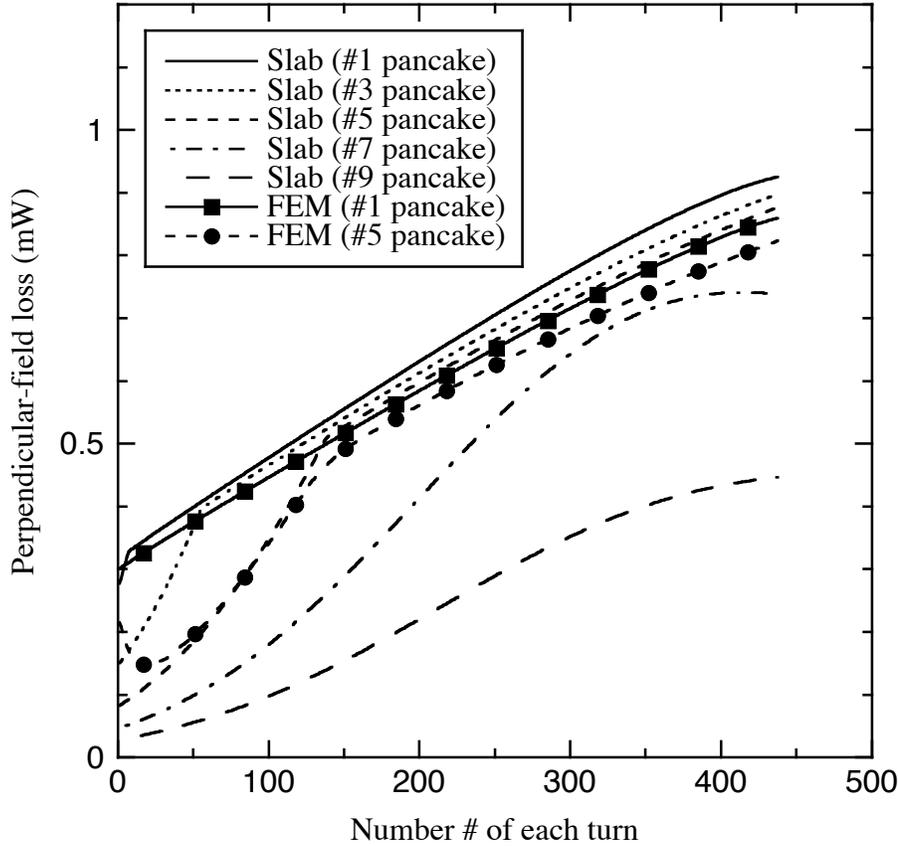}
\caption{Profiles of perpendicular-field losses in selected pancake coils of HTS insert.
The curves without symbols are obtained using the slab approximation, whereas the curves with symbols represent the numerical results on the basis of the finite element method.}
\label{fig7}
\end{figure}

\Fref{fig7} shows the profiles of perpendicular-field losses at $t=60$~min in the selected pancake coils.
The curves without symbols are obtained using the slab approximation, whereas the curves with symbols represent the numerical results obtained using the FEM.
Turns \#1 and \#438 are located at the innermost and outermost parts of the pancake coil, respectively.
It can be seen that there are kinks in the profiles for pancakes \#1, \#3, and \#5, and therefore the segments of the turns outside the kinks are exposed to perpendicular magnetic fields larger than the full penetration fields.
In the cases of pancakes \#7 and \#9, on the other hand, the perpendicular fields applied to all the turns are smaller than the full penetration fields.
It is also found that both the profiles from the slab approximation and the FEM show reasonably good agreement, but there are small discrepancies between them.
One of the discrepancies is that the end effect can be seen for pancake \#5 in the FEM below the full penetration field, as discussed in \sref{sec:magnetic_interaction}.
On the other hand, there is no end effect above the full penetration field.
These results indicate that the perpendicular-field losses can be estimated with relatively small errors using the slab approximation.

\begin{figure}[tbp]
\centering
\includegraphics*[scale=1.0]{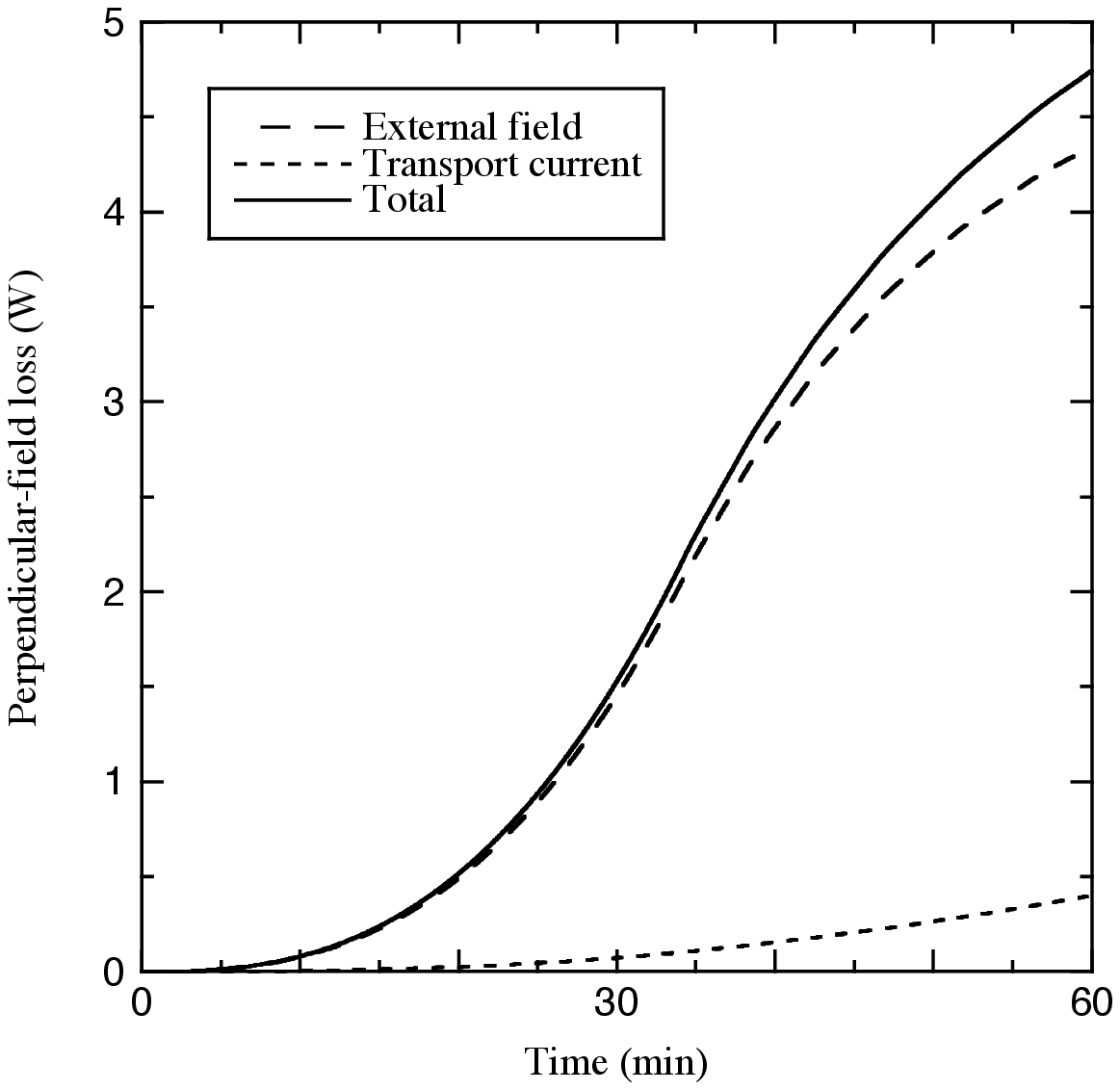}
\caption{Influence of transport current on perpendicular-field losses in HTS insert estimated using slab approximation.}
\label{fig8}
\end{figure}

\Fref{fig8} shows the influence of the transport current on the perpendicular-field losses in all the pancake coils of the HTS insert, estimated using the slab approximation.
In order to draw \fref{fig8}, the theoretical expressions \eref{eqn:loss4} are divided into two parts: the contributions from the external magnetic field and from the transport current.
It can be seen that the influence of the transport current on the perpendicular-field losses is significant and cannot be ignored. It is also found that the total perpendicular-field losses increase monotonically with time.

\begin{figure}[tbp]
\centering
\includegraphics*[scale=1.0]{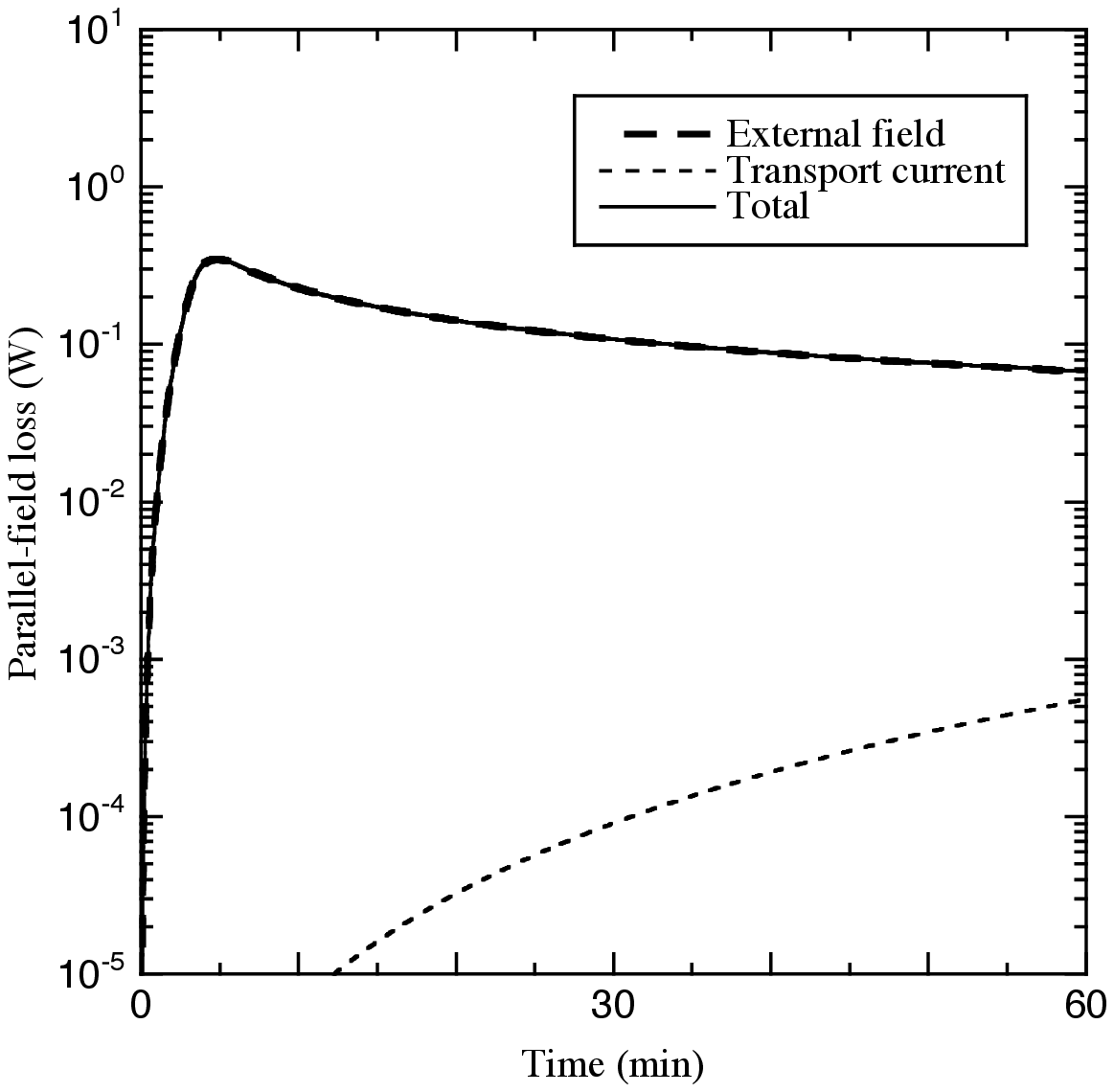}
\caption{Influence of transport current on parallel-field losses in HTS insert estimated using slab approximation.}
\label{fig9}
\end{figure}

\Fref{fig9} shows the numerical results for the AC losses during energization of the axial components, $B_z$, of the magnetic fields parallel to the HTS tapes.
These are determined for all the pancake coils of the HTS insert, estimated using the theoretical expressions \eref{eqn:loss4} for an infinite slab with width equal to the thickness $d$ of the SC layer.
It can be seen that the influence of the transport current on the parallel-field losses is negligible.
It is also found that the parallel-field losses are maximized in about 5~min, and, on average, the parallel fields applied to all the turns reach the full penetration fields.
Subsequently, the parallel-field losses decrease due to the reduction of the critical current density.

\begin{figure}[tbp]
\centering
\includegraphics*[scale=1.0]{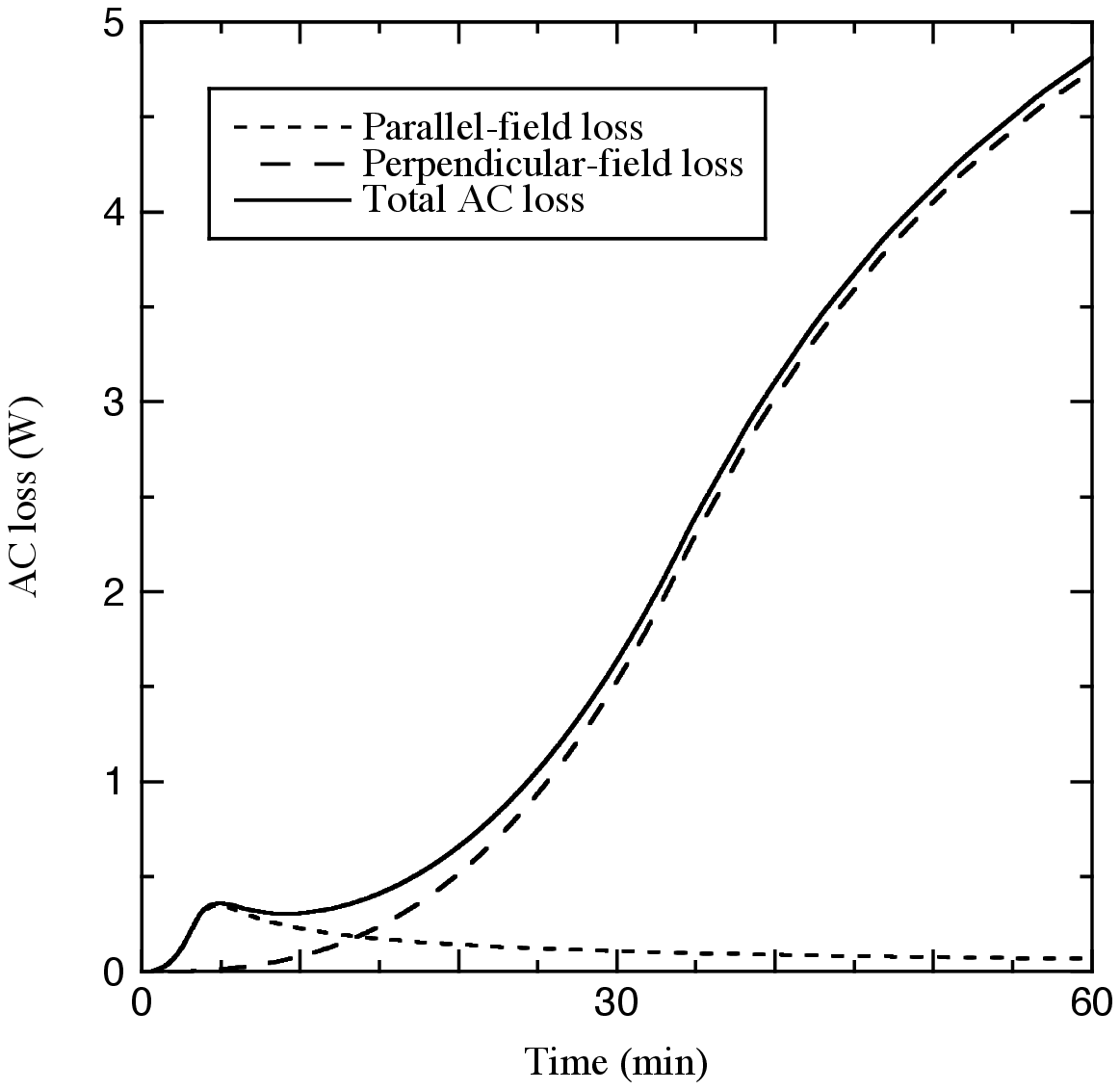}
\caption{Time evolution of total AC losses in HTS insert estimated using slab approximation.}
\label{fig10}
\end{figure}

\Fref{fig10} shows the time evolution of the total AC losses in the HTS insert, estimated using the slab approximation.
The perpendicular- and parallel-field losses in figures~\ref{fig8} and \ref{fig9} are summed simply to obtain the total AC loss on the basis of \eref{eqn:loss3}.
When the HTS insert is first energized, the parallel-field losses are dominant.
Subsequently, the perpendicular-field losses become dominant, and the total losses increase almost monotonically.
The wattage of about 5~W at $t=60$~min is slightly larger than the cooling power of 3~W at 4.2~K for the cryocoolers under consideration.
This means that the operating temperature of the HTS insert would be balanced at a slightly higher temperature due to the drastic improvement in cooling power that the cryocoolers possess at higher temperatures.

\section{Conclusions}

The AC losses in the HTS insert coil for a high field magnet during energization have been numerically estimated.
The slab approximation can be used to calculate not only the parallel-field losses due to the axial components of the local magnetic fields, but also the perpendicular-field losses due to their radial components.
Further investigations, such as an estimation of AC loss combined with thermal analysis and confirmation of this approach by experiment, will be required.


\section*{References}
\begin{thebibliography}{99}
\bibitem{Amemiya} Amemiya N, Jiang Z, Iijima Y, Kakimoto K and Saitoh T 2004 \SUST {\bf 17} 983
\bibitem{Mawatari07} Mawatari Y and Kajikawa K 2007 {\it Appl. Phys. Lett.} {\bf 90} 022506
\bibitem{Kajikawa_PC10} Kajikawa K, Funaki K, Shikimachi K, Hirano N and Nagaya S 2010 {\it Physica} C {\bf 470} 1321
\bibitem{Gavrilin} Gavrilin A V, Lu J, Bai H, Hilton D K, Markiewicz W D and Weijers H W 2013 {\it IEEE Trans. Appl. Supercond.} {\bf 23} 4300704
\bibitem{Lu} Lu J, Abraimov D V, Polyanskii A A, Gavrilin A V, Hilton D K, Markiewicz W D and Weijers H W 2013 {\it IEEE Trans. Appl. Supercond.} {\bf 23} 8200804
\bibitem{Zenkevitch} Zenkevitch V B, Zheltov V V and Romanyuk A S 1978 {\it Cryogenics} {\bf 18} 93
\bibitem{Sumiyoshi} Sumiyoshi F, Irie F and Yoshida K 1980 \JAP {\bf 51} 3807
\bibitem{Kajikawa_IPCS167} Kajikawa K, Takenaka A, Iwakuma M and Funaki K 2000 {\it Inst. Phys. Conf. Ser.} {\bf 167} 931
\bibitem{Suenaga} Suenaga M, Chiba T, Ashworth S P, Welch D O and Holesinger T G 2000 \JAP {\bf 88} 2709
\bibitem{Kajikawa_IEEJ01} Kajikawa K, Nishimura M, Moriyama H, Iwakuma M and Funaki K 2001 {\it Trans. IEE Jpn.} {\bf 121-B} 1283; Kajikawa K, Nishimura M, Moriyama H, Iwakuma M and Funaki K 2002 {\it Electr. Eng. Jpn.} {\bf 141} 50
\bibitem{Iwakuma} Iwakuma M, Toyota K, Nigo M, Kiss T, Funaki K, Iijima Y, Saitoh T, Yamada Y and Shiohara Y 2004 {\it Physica} C {\bf 412--414} 983
\bibitem{Grilli06} Grilli F, Ashworth S P and Stavrev S 2006 {\it Physica} C {\bf 434} 185
\bibitem{Yuan} Yuan W, Campbell A M and Coombs T A 2009 \SUST {\bf 22} 075028
\bibitem{Kajikawa_PC09} Kajikawa K, Funaki K, Shikimachi K, Hirano N and Nagaya S 2009 {\it Physica} C {\bf 469} 1436
\bibitem{Prigozhin} Prigozhin L and Sokolovsky V 2011 \SUST {\bf 24} 075012
\bibitem{Grilli07} Grilli F and Ashworth S P 2007 \SUST {\bf 20} 794
\bibitem{Pardo} Pardo E 2008 \SUST {\bf 21} 065014
\bibitem{Zermeno} Zerme\~{n}o V M R and Grilli F 2014 \SUST {\bf 27} 044025
\bibitem{Sato} Sato S and Amemiya N 2006 {\it IEEE Trans. Appl. Supercond.} {\bf 16} 127
\bibitem{Kajikawa_PC06} Kajikawa K, Mawatari Y, Iiyama Y, Hayashi T, Enpuku K, Funaki K, Furuse M and Fuchino S 2006 {\it Physica} C {\bf 445--448} 1058
\bibitem{Awaji_IEEE-TAS} Awaji S, Watanabe K, Oguro H, Hanai S, Miyazaki H, Takahashi M, Ioka S, Sugimoto M, Tsubouchi H, Fujita S, Daibo M, Iijima Y and Kumakura H 2014 {\it IEEE Trans. Appl. Supercond.} {\bf 24} 4302005
\bibitem{Fukuda} Fukuda Y, Toyota K, Kajikawa K, Iwakuma M and Funaki K 2003 {\it IEEE Trans. Appl. Supercond.} {\bf 13} 3610
\bibitem{Awaji_CSSJ} Awaji S, Watanabe K, Oguro H, Mitose T, Kajikawa K, Fujita S, Daibo M, Iijima Y, Miyazaki H, Takahashi M and Ioka S 2013 {\it Abst. Cryo. Supercond. Soc. Jpn. Conf.} {\bf 88} 12
\bibitem{Hashizume} Hashizume H and Miya K 1989 {\it Fusion Eng. Des.} {\bf 7} 293
\bibitem{Ichiki} Ichiki Y and Ohsaki H 2004 {\it Physica} C {\bf 412--414} 1015
\bibitem{Bean} Bean C P 1962 \PRL {\bf 8} 250
\bibitem{Kajikawa_IEEE-TAS03} Kajikawa K, Hayashi T, Yoshida R, Iwakuma M and Funaki K 2003 {\it IEEE Trans. Appl. Supercond.} {\bf 13} 3630
\bibitem{Kiss} Kiss T and Okamoto H 2001 {\it IEEE Trans. Appl. Supercond.} {\bf 11} 3900
\bibitem{Halse} Halse M R 1970 \JPD {\bf 3} 717
\bibitem{Brandt} Brandt E H and Indenbom M 1993 \PR B {\bf 48} 12893
\bibitem{Zeldov} Zeldov E, Clem J R, McElfresh M and Darwin M 1994 \PR B {\bf 49} 9802
\bibitem{Mawatari96} Mawatari Y 1996 \PR B {\bf 54} 13215
\bibitem{London} London H 1963 \PL {\bf 6} 162
\bibitem{Clem} Clem J R, Claassen J H and Mawatari Y 2007 \SUST {\bf 20} 1130
\end{thebibliography}
\end{document}